# Pluggable AOP — Designing Aspect Mechanisms for Third-party Composition[*]

Sergei Kojarski   David H. Lorenz
Northeastern University
Boston, Massachusetts 02115 USA
{kojarski,lorenz}@ccs.neu.edu

## ABSTRACT
Studies of Aspect-Oriented Programming (AOP) usually focus on a language in which a specific aspect extension is integrated with a base language. Languages specified in this manner have a fixed, non-extensible AOP functionality. In this paper we consider the more general case of integrating a base language with a set of domain specific third-party aspect extensions for that language. We present a general mixin-based method for implementing aspect extensions in such a way that multiple, independently developed, dynamic aspect extensions can be subject to third-party composition and work collaboratively.

## 1. INTRODUCTION

A current trend in Aspect-Oriented Programming (AOP [26]) is the usage of general-purpose AOP languages (AOPLs). However, a general-purpose AOPL lacks the expressiveness to tackle all cases of crosscutting. A solution to unanticipated crosscutting concerns is to create and combine different domain-specific aspect extensions to form new AOP functionality [42]. As of yet, there is no methodology to facilitate this process.

Studies of AOP typically consider the semantics for an AOPL that integrates a certain aspect extension, $Ext_1$, with a base language, Base. For example, $Ext_1$ might be (a simplified version of) AspectJ [25] and Base (a simplified version of) Java [3]. The semantics for the integration Base $\times$ $Ext_1$ is achieved by amending the semantics for the base language. Given a pair of programs $\langle base, aspect_1 \rangle \in$ Base $\times$ $Ext_1$, the amended semantics explain the meaning of $base$ in the presence of $aspect_1$.

Unfortunately, the semantics for the aspect extension and that for the base language become tangled in the process of integration. Consequently, it is difficult to reuse or combine aspect extensions. For each newly introduced aspect extension, say $Ext_2$, the semantics for Base $\times$ $Ext_2$ needs to be reworked. Moreover, given the semantics for Base $\times$ $Ext_1$ and the semantics for Base $\times$ $Ext_2$, the semantics for Base $\times$ $Ext_1$ $\times$ $Ext_2$ is undefined even though $Ext_1$ and $Ext_2$ are both aspect extensions to the same base language.

In this paper we resolve this difficulty by considering a more general question:

THE ASPECT EXTENSION COMPOSITION QUESTION: *Given a base language, Base, and a set $\{Ext_1, \ldots, Ext_n\}$ of independent aspect extensions to Base, what is the meaning of a program $base \in$ Base in the base language in the presence of $n$ aspect programs $\langle aspect_1, \ldots, aspect_n \rangle \in$ $Ext_1 \times \cdots \times Ext_n$ written in the $n$ different aspect extensions?*

Ability to compose distinct aspect extensions is of great practical importance (Section 2). Addressing the general composition question also provides in the special case where $n = 1$ a better encapsulation of the semantics for a single aspect extension.

### 1.1 Combining Two Aspect Extensions

Answering the aspect extension composition question is difficult even for $n = 2$. Let MyBase be a procedural language, and consider two independent, third-party aspect extensions to MyBase. The first, $HisExt_1$, capable of intercepting procedure calls and similar in flavor to AspectJ. The other, $HerExt_2$, an aspect extension to MyBase capable of intercepting calls to the primitive division operator for catching a division by zero before it even happens (as opposed to catching a division by zero exception after it occurs), a capability that AspectJ lacks.[1] Both call interception (e.g., [27]) and checking if a divisor is zero (e.g., [5, 28, 18]) are benchmarks often used in connection with aspects.

W.l.o.g., assume $HisExt_1$ is created before $HerExt_2$ is even conceived. If $HisExt_1$ is to eventually work collaboratively with another aspect extension, e.g., $HerExt_2$, the implementation of HisExt$_1$ must take special care to expose its AOP effect, *and only its* effect, in terms of MyBase. This is because an $aspect_2$ program written in $HerExt_2$ would need to intercept divisions by zero not only in the base program $base$ but also in advice introduced by an $aspect_1$ program written in $HisExt_1$.

Failing to reify a division by zero in $aspect_1$ might cause a false-negative effect in $HerExt_2$. Meanwhile, $aspect_2$ must not intercept divisions by zero, if any, in the implementation mechanism of either $HisExt_1$ or $HerExt_2$. Reifying a division by zero in the implementation mechanism might cause a false-positive effect in $HerExt_2$.

---

[*]This research was supported in part by the National Science Foundation (NSF) Science of Design program under Grant Number CCF-0438971, and by the Institute for Complex Scientific Software at Northeastern University (http://www.icss.neu.edu).

[1]AspectJ can neither advise primitives nor arguments.

Similarly, $aspect_1$ must intercept not only procedure calls in $base$ but also any matching procedure call introduced by $aspect_2$. $aspect_1$ must not, however, intercept internal procedure calls that are a part of the implementation mechanism of either $\mathsf{HisExt}_1$ or $\mathsf{HerExt}_2$.

Note that generally aspect extensions present incompatible levels of AOP granularity [30]. In our example, $aspect_1$ is not expressible in $\mathsf{HerExt}_2$, and $aspect_2$ is not expressible in $\mathsf{HisExt}_1$. Therefore the problem of integrating the two cannot be reduced to translating $aspect_1$ to $\mathsf{HerExt}_2$ or translating $aspect_2$ to $\mathsf{HisExt}_1$ and using just one aspect extension. This distinguishes our objective from the purpose of frameworks (like XAspects [38]) that rely on the use of a general purpose AOPL (like AspectJ).

In the sequel, a *base mechanism* denotes an implementation of a base language semantics, an *aspect mechanism* denotes an implementation of an aspect extension semantics, and a *multi mechanism* denotes an implementation of a multi-extension AOPL.

## 1.2 Objective and Contribution

We describe a general method for implementing the base mechanism and the aspect mechanisms in such a way that multiple, independent aspect mechanisms can be subject to third-party composition and work collaboratively. By third-party composition of aspect mechanisms we mean a semantical framework in which distinct aspect mechanisms can be assembled with the base mechanism into a meaningful multi mechanism without modifying the individual mechanisms. The mechanisms are said to be collaborative units of composition if the semantics of the composed multi mechanism can be derived from the semantics of the mechanisms that comprise it.

More precisely, let $\mathcal{B}$ denote the base mechanism for $\mathsf{Base}$. Let $\mathcal{M}_1, \ldots, \mathcal{M}_n$ denote the aspect mechanisms for $\mathsf{Ext}_1, \ldots, \mathsf{Ext}_n$, respectively. The *aspect mechanism composition problem* is to enable the third-party composition of $\mathcal{M}_1, \ldots, \mathcal{M}_n$ with $\mathcal{B}$ into a multi mechanism $\mathcal{A}$, in a mannar similar to the assembly of software components:[2]

- **Units of independent production.** The aspect mechanisms $\mathcal{M}_1, \ldots, \mathcal{M}_n$ are independently defined. The base mechanism $\mathcal{B}$ is defined independently from $\mathcal{M}_1, \ldots, \mathcal{M}_n$. To enable the composition, $\mathcal{M}_1, \ldots, \mathcal{M}_n$ rely only on $\mathcal{B}$ and have an explicit context dependency only on $\mathcal{A}$.

- **Units of composition.** The mechanisms are subject to third-party composition. The multi mechanism $\mathcal{A}$ for the combined AOP language is constructed (denoted by a $\boxplus$ combinator) by composing the base mechanism with the aspect mechanisms without altering them: $\mathcal{A} = \boxplus \langle \mathcal{B}, \mathcal{M}_1, \ldots, \mathcal{M}_n \rangle$.

- **Units of collaboration.** The semantics for the composed multi mechanism $\mathcal{A}$ is the "sum" of the semantics provided by all the mechanisms.

Independence enables third-party development of aspect mechanisms; composability enables third-party composition of aspect mechanisms; and collaboration enables the desired behavior in the constructed AOP language.

---
[2] A software component is a unit of composition with contractually specified interfaces and explicit context dependencies only. A software component can be deployed independently and is subject to third-party composition [40].

Specifically, our approach enables third-party composition of dynamic aspect mechanisms. We illustrate our solution for expression evaluation semantics. We model each aspect mechanism as a transformation function that revises the evaluation semantics for expressions.

## 1.3 Outline

In the rest of this paper, we demonstrate our solution to the aspect mechanism composition problem concretely through the implementation of interpreters. The next section motivates the need for composing multiple aspect extensions and demonstrates the lack of integration support in current aspect mechanisms. Section 3 presents a concrete instance of the problem: a base language My-Base with two aspect extensions, $\mathsf{HisExt}_1$ and $\mathsf{HerExt}_2$. We present their syntax and analyze a runnable programming example implemented in our framework. In Section 4 we present our approach for the general case of integrating $n$ aspect mechanisms. In Section 5 we revisit the example shown in Section 3 and formally demonstrate our approach by constructing the semantics for My-Base, $\mathsf{HisExt}_1$, and $\mathsf{HerExt}_2$.

## 2. MOTIVATION

There is a growing need for the simultaneous use of multiple domain-specific aspect extensions. The need steams mainly from the favorable trade-offs that a domain-specific aspect extension can offer over a general purpose AOPL:

- *Abstraction.* A general purpose AOPL offers low-level abstractions for covering a wide range of crosscutting concerns. Domain specific aspect extensions, in contrast, can offer abstractions more appropriate for the crosscutting cases in the domain at hand, letting the programmer concentrate on the problem, rather then on low-level details.

- *Granularity.* The granularity of an aspect extension dictates all possible concern effect points within an application. Combining domain-specific aspect extensions allows to overcome the fixed granularity limitation of general purpose AOPLs [30].

- *Expressiveness versus Complexity.* The granularity of a general-purpose AOPL exposes a non-linear relationship between the language expressiveness and complexity. An increase in the language granularity would significantly increase the language complexity while achieving a relatively small increase in expressiveness. Domain specific aspect extensions, in contrast, can offer independent diverse ontologies [48].

The need also arises from the sheer abundance of available aspect extensions (and their evolving aspect libraries). For the Java programming language alone there are numerous aspect extensions that are being used in a variety of commercial and research projects. These include: AspectJ (ajc [12] and abc [4]), AspectWerkz [6], COOL [29], JBoss-AOP [2], JAsCo [43], Object Teams [21], ComposeJ [50], to name just a few.[3] Ability to use these aspect extensions together will allow to reuse exiting (and future) aspect libraries written for the different aspect extensions.

Unfortunately, little support is provided for the integration of distinct aspect mechanisms. Each aspect mechanism creates its own

---
[3] For a complete list of commercial and research aspect extensions see http://www.aosd.net/technology/.

Listing 1: A non-synchronized bounded buffer

```java
public class BoundedBuffer {

  private Object[] buffer;
  private int usedSlots = 0;
  private int writePos = 0;
  private int readPos = 0;
  private static BoundedBuffer singltn = null;

  public static BoundedBuffer getInstance() {
    return singltn;
  }

  public BoundedBuffer (int capacity) {
    this.buffer = new Object[capacity];
    singltn = this;
  }

  public Object remove() {
    if (usedSlots == 0) {return null;}
    Object result = buffer[readPos];
    buffer[readPos] = null;
    usedSlots--; readPos++;
    if (readPos==buffer.length) readPos=0;
    return result;
  }

  public void add(Object obj) throws Exception {
    if (usedSlots==buffer.length)
      throw new Exception("buffer is full");
    buffer[writePos] = obj;
    usedSlots++;
    writePos++;
    if (writePos==buffer.length) writePos=0;
  }
}
```

Listing 2: Synchronization aspect in COOL

```
coordinator BoundedBuffer {
  selfex {add, remove},
  mutex {add, remove};
}
```

Listing 3: Synchronization aspect in AspectJ

```java
public aspect BufferSyncAspect {
  private Object remove_thread=null;
  private Object add_thread=null;

  Object around():
    execution(Object BoundedBuffer.remove()) {
    Object this_thread = Thread.currentThread();
    synchronized(this) {
      while ((remove_thread!=null &&
          remove_thread!=this_thread) ||
         (add_thread!=null &&
          add_thread!=this_thread))
        try {wait();
        } catch (InterruptedException e) {}
      remove_thread = this_thread;
    }
    Object result = proceed();
    synchronized(this) {
      remove_thread = null;
      notifyAll();
    }
    return result;
  }

  void around() throws Exception:
    execution(void BoundedBuffer.add(Object)) {
    Object this_thread = Thread.currentThread();
    synchronized(this) {
      while ((remove_thread!=null &&
          remove_thread!=this_thread) ||
         (add_thread!=null &&
          add_thread!=this_thread))
        try {wait();
        } catch (InterruptedException e) {}
      add_thread = this_thread;
    }
    try{proceed();}
    finally {
      synchronized(this) {
        add_thread = null;
        notifyAll();
      }
    }
  }
}
```

unique program representation which often excludes foreign aspects. Consequently, interaction between multiple aspect mechanisms operating on a single program can produce unexpected or incoherent results.

## 2.1 Example

Consider a bounded buffer example implemented in Java (Listing 1). Suppose you have three aspect extensions to Java at your disposal:

- COOL [29]—a domain-specific aspect extension for expressing coordination of threads;
- AspectWerkz [6]—a general purpose lightweight AOP framework for Java;
- AspectJ—a general purpose aspect extension for Java;

and two concerns to address, namely, a synchronization concern and a tracing concern.

### 2.1.1 COOL versus AspectJ

The synchronization concern can be expressed as a coordinator aspect in COOL (e.g., Listing 2) or alternatively as an aspect in AspectJ (e.g., Listing 3).

The COOL aspect (Listing 2) provides an elegant declarative description of the desired synchronization. The **mutex** exclusion set {add, remove} specifies that add may not be executed by a thread while remove is being executed by a different thread, and vice versa. In addition, the **selfex** exclusion set prohibits different threads from simultaneously executing either add or remove.[4]

The COOL code is expressive, concise, readable, and easy to understand. It provides the right abstractions. Studies [33, 46, 32, 47] have shown that "participants could look at COOL code and understand its effect without having to analyze vast parts of the rest of the code", and that "COOL as a synchronization aspect language eased the debugging of multi-threaded programs, compared to the

---
[4]However, the same thread is not prohibited from entering both add and remove.

Listing 4: Logger aspect in AspectWerkz

```
/** @Aspect("perJVM") */
public class AWLogger {
  /**@Before call(* *.*(..))&&!cflow(within(AWLogger))*/
  public void log(JoinPoint jp) {
    System.out.println("AW:"+jp.getSignature());
  }
}
```

Listing 5: Buffer Logger

```
public aspect BufferLogger {

  pointcut toLog():
    call(* *.*(..)) && !cflow(within(BufferLogger)
        );

  before(): toLog() {
    log("ENTER",thisJoinPoint);
  }
  after() returning: toLog() {
    log("EXIT",thisJoinPoint);
  }
  after() throwing: toLog() {
    log("THROW",thisJoinPoint);
  }

  protected void log(String aType,JoinPoint jp) {
    BoundedBuffer buf=BoundedBuffer.getInstance();
    if (buf==null) return;
    try{buf.add(jp);} catch (Exception e) {
      System.out.println(e.getMessage());
    }
  }
}
```

ability to debug the same program written in Java" [45].

While it is possible to express the same concern in AspectJ, the code will be much longer. In comparison to the COOL code, the AspectJ implementation (Listing 3) requires 10 times more lines of code. It is also harder to explain. The aspect implements a monitor using two condition variables `remove_thread` and `add_thread`. Using two pieces of **around execution** advice, the aspect obtains locks (`remove_thread` and `add_thread`) for the duration of executing **proceed** (execution of `remove` and `add`, respectively). This guarentees that no more than one thread operates on the buffer at a time. If `remove_thread` or `add_thread` are locked by some other thread, the advice waits. When the thread has a lock, it runs **proceed** and afterwards releases the lock by signaling `notifyAll()`, which in turn wakes up other waiting threads.

### 2.1.2 AspectWerkz + AspectJ

Semantically, the underlying mechanisms of AspectWerkz and AspectJ are essentially equivalent. Yet, their syntactical differences present programmers with a desired choice of alternatives. Recently is was announced that AspectWerkz has joined the AspectJ project to bring the key features of AspectWerkz to the AspectJ 5 platform [7]. This merger will allow aspects like those in Listing 4 and Listing 5 to run side by side.

Listing 4 is a simple tracing aspect in AspectWerkz. The code is plain Java. The annotation `@Aspect("perJVM")` specifies that the `AWLogger` class is actually a singleton aspect. The annotation `@Before call(* *.*(..)) && !cflow(within(AWLogger))` specifies that the `log` method is to be called for every method call not in the dynamic control flow of methods in `AWLogger`.

Listing 5 is an auditing aspect in AspectJ. The `toLog()` pointcut specifies that every method call should be logged. The **before**, **after**()**returning**, and **after**()**throwing** advice add log messages to the buffer.

Arguably, if AspectWerkz and AspectJ were designed to be composable third-party aspect mechanisms, building AspectJ 5 would have been much easier. Moreover, third-party composition of aspect mechanisms would have made other domain specific combinations possible, like combining COOL with AspectWerkz and Java.

## 2.2 Lack of Integration Support

Unfortunately, current aspect mechanisms fail to compose correctly. We demonstrate this failure on the bounded buffer example for two commonly used approaches:

- *Translation.* Aspect programs in different aspect extensions can be translated to a common target aspect extension.

- *Instrumentation.* Aspect mechanisms can be implemented by means of program instrumentation. Such multiple independent aspect mechanisms can be trivially composed by passing the output of one aspect mechanism as the input to another aspect mechanism.

### 2.2.1 No Behavior-Preserving Translation

The translation approach requires the expressiveness of the target aspect extension to support arbitrary granularity. Even when granularity does not pose a problem, a translation from one aspect language to another will not generally preserve the behavior of the source aspect program in the presence of other aspects. Consider the synchronization concern implementation in COOL (Listing 2). Translating it to AspectJ (Listing 3) results in an aspect that seems to be a correct substitution for the COOL coordination aspect, but in the presence of the Logger aspect (Listing 5) is actually not.

A property of the COOL synchronization concern is transparency with respect to the AspectJ logging concerns. There should not be any interference between the two. The COOL aspect does not contain any join points that should be visible to the AspectJ mechanism. This property is not preserved in the translation. Calls to `wait` (Listing 3, lines 13 and 33) and `notifyAll` (Listing 3, lines 20 and 41), which do not exist in the COOL code, will nonetheless be unexpectedly reflected by the logger.[5]

Worse yet, the unexpected join points in the target program may break existing invariants, resulting in our case in a deadlock. An implicit invariant of the COOL aspect is that if both `add` and `remove` are not currently executing by some other thread, then the thread can enter and execute them. The AspectJ synchronization aspect, however, violates this invariant. Assume that two threads concurrently access the buffer. The first thread acquires the lock, while the second invokes `wait` on the `BufferSyncAspect` object. However, before `wait` is invoked, the `BufferLogger` aspect calls `BoundedBuffer.add` (Listing 5, line 19). The latter call causes the second thread to enter the guarded code *again* and trigger a *second* call to `wait`.[6] Since the second `wait` call is in the **`cflow`** of the logger, it is not advised, and the thread finally suspends. When the first thread releases the lock, the second thread wakes up after the *second* `wait`. It acquires the lock, completes the advice execution, releases the lock, and proceeds to the *first* `wait` invocation. At this point, the buffer is not locked; the second thread waits on the `BufferSyncAspect` object monitor; and if no other thread ever accesses the buffer, the second thread waits for ever—-deadlock!

---
[5]Note that calls to `wait` and `notifyAll` cannot be avoided.
[6]Assuming that the first thread still owns the lock.

### 2.2.2 *No Correct Order for Sequential Processing*
One would expect the two aspects written in AspectWerkz (Listing 4) and AspectJ (Listing 5) to interact as if they were two aspects written in a single aspect extension (e.g., the future AspectJ 5 platform). On the one hand, the AspectJ logger should log all method calls within the `AWLogger` aspect. On the other hand, the AspectWerkz logger should log all method calls within `BufferLogger`. (And both should log all method calls in the base program as well.)

However, applying the AspectJ and AspectWerkz instrumentation mechanisms sequentially, in any order, produces an unexpected result. The mechanism that is run first may not be able to interpret the second extension's aspect program. Specifically, the AspectWerkz mechanism does not understand AspectJ's syntax. It can be applied to the bounded buffer code but not to the `BufferLogger` aspect. Thus, when AspectWerkz is run first, some expected log messages will be missing.

The mechanism that is run last logs method calls that are not supposed to be logged. For example, when AspectWerkz is run second, the following unexpected log message is generated by the `AWLogger` aspect:

```
AW:public void BufferLogger.
    ajc$afterReturning$BufferLogger$2$ba1fbd8a(
    org.aspectj.lang.JoinPoint)
```

## 3. PROBLEM INSTANCE
We now return to MyBase, HisExt$_1$, and HerExt$_2$ in order to analyze the problem and illustrate our approach concretely. After a brief introduction to the syntax, we informally explain MyBase, HisExt$_1$, and HerExt$_2$ through a programming example. The code fragments are actual running code in our implementation, and their semantics is formally presented in Section 5.

### 3.1 Syntax

#### 3.1.1 MyBase *Syntax*
The syntax of MyBase is given in Figure 1. MyBase is a procedural language. Procedures are mutually-recursive with call-by-value semantics. The set of procedures is immutable at run-time. Expressed values are either booleans or numbers (but not procedures). The execution of a program starts by evaluating the body of a procedure named **`main`**.

#### 3.1.2 HisExt$_1$ *Syntax*
The syntax for HisExt$_1$ is given in Figure 2. HisExt$_1$ is a simple AspectJ-like aspect extension to MyBase. HisExt$_1$ allows one to impose advice around procedure calls and procedure executions. Advice code is declared in a manner similar to procedures. Like in AspectJ, the set of advice is immutable at run-time. Each advice has two parts: a pointcut designator and an advice body expression. Atomic pointcuts are `pcall-pcd`, `pexecution-pcd`, `cflow-pcd`, and `args-pcd`. The `and-pcd` and `or-pcd` allows one to combine several pointcuts under conjunction and disjunction, respectively. Unlike AspectJ, `around` is the only advice kind in HisExt$_1$. There is no support for patterns in pointcut designators. HisExt$_1$ introduces a new `proceed-exp` expression, which is valid only within an HisExt$_1$ advice body expression.

#### 3.1.3 HerExt$_2$ *Syntax*
HerExt$_2$ allows one to declare a set of exception handlers in MyBase for catching and handling division by zero before an exception occurs. Advice code in HerExt$_2$ specifies an exception handler expression. A guard clause allows one to specify a dynamic scope for the handler. HerExt$_2$ introduces a new expression, namely `raise-exp`, which is allowed within a handler. It passes the exception handling to the next handler (in a manner, similar to `proceed-exp` of HisExt$_1$). The syntax of the language is given in Figure 3.

The semantics for HerExt$_2$ is straightforward. Whenever the second argument to the division primitive evaluates to zero, the advice handler (if one exits) is invoked. The handler is evaluated and the result value substitutes the offending zero in the second argument to the division primitive, and the program execution resumes.

Listing 8 shows an aspect we can write in HerExt$_2$. This aspect resumes the execution with the value of `Precision(1)` whenever the second argument of a division primitive evaluates to 0 within the control flow of the `SQRT` procedure.

### 3.2 A Programming Example
The semantics for the base procedural language MyBase and the aspect extensions HisExt$_1$ and HerExt$_2$ are implemented as interpreters [19]. The example presented here is a simple executable arithmetic program in MyBase for computing the square root of a given number. While simple, the example is representative in terms of illustrating the complexity of achieving collaboration among aspect extensions, and its semantics serves as a proof of concept.

The procedure `SQRT` in Listing 6 implements in MyBase a simple approximation algorithm using a sequence generated by a recurrence relation:

$a_0=approximation$ ; **repeat** $a_n=f(a_{n-1})$ **until** $precise$

By default, it sets $a_0 = 0$, and calls `SqrtIter` to generate the recurrence sequence:

$$a_n = a_{n-1} + \epsilon$$

until $(a_n)^2 > x$. The procedure `Improve` generates the next element in the sequence; `IsPreciseEnough?` checks the termination condition; and the value $\epsilon = \epsilon(x)$ is computed as a function of $x$ by the procedure `Precision`.

The resulted computation of $\sqrt{x}$ is inaccurate and extremely inefficient. However, it serves our purpose well. We will non-intrusively improve its efficiency using an aspect in HisExt$_1$. We will correct its behavior for the singular point $x = 0$ using HerExt$_2$.

The code in Listing 7, written in HisExt$_1$, advises the base code for drastically improving its efficiency and accuracy. Four pieces of advice are used. The first around advice (lines 202–204) intercepts executions of the procedure `Improve` and instead applies

```
Program         ::=  Declaration                                  Program
Declaration     ::=  "program" "{" Procedure* "}"                 Declaration
Procedure       ::=  "procedure" PName "(" Id* ")" Exps            Procedure
Exps            ::=  lit-exp | true-exp | false-exp |
                     var-exp | app-exp | begin-exp | if-exp |
                     assign-exp | let-exp | primapp-exp            Expressions
lit-exp         ::=  Number                                        Numbers
true-exp        ::=  "true"                                        True
false-exp       ::=  "false"                                       False
var-exp         ::=  Id                                            Id meaning
app-exp         ::=  "call" PName "(" Exps* ")"                    Procedure call
begin-exp       ::=  "{" Exps ( ";" Exps )* "}"                    Block
if-exp          ::=  "if" Exps "then" Exps "else" Exps             Conditional
assign-exp      ::=  "set" Id "=" Exps                             Assignment
let-exp         ::=  "let" ( Id "=" Exps )* "in" Exps              Let
primapp-exp     ::=  Prim "(" Exps* ")"                            Primitive application
Prim            ::=  "+" | "-" | "*" | "/"                         Primitives
Id                                                                 Identifier
PName                                                              Procedure name
Number                                                             Numbers
```

Figure 1: MyBase syntax

```
AOP1-Program      ::=  AOP1-Declaration                            HisExt_1 program
AOP1-Declaration  ::=  "aop1" "{" Advice* "}"                      HisExt_1 declaration
Advice            ::=  "around" ":" Pointcut Exps_1                Advice
Pointcut          ::=  call-pcd | exec-pcd | cflow-pcd |
                       args-pcd | and-pcd | or-pcd                 Pointcut designators
call-pcd          ::=  "pcall" "(" PName ")"                       Procedure call pcd
exec-pcd          ::=  "pexecution" "(" PName ")"                  Procedure execution pcd
cflow-pcd         ::=  "cflow" "(" PName ")"                       Control flow pcd
args-pcd          ::=  "args" "(" Id* ")"                          Argument pcd
and-pcd           ::=  "and" "(" Pointcut* ")"                     Conjunction pcd
or-pcd            ::=  "or" "(" Pointcut* ")"                      Disjunction pcd
Exps_1            ::=  Exps | proceed-exp                          Advice expressions
proceed-exp       ::=  "proceed"                                   Proceed exp
```

Figure 2: HisExt$_1$ syntax

```
AOP2-Program      ::=  AOP2-Declaration                            HerExt_2 program
AOP2-Declaration  ::=  "aop2" "{" Handler* "}"                     HerExt_2 declaration
Handler           ::=  "guard_cflow" PName "resume_with" Exps_2    Handlers
Exps_2            ::=  Exps | raise-exp                            Handler expressions
raise-exp         ::=  "raise"                                     Raise expressions
```

Figure 3: HerExt$_2$ syntax

Listing 6: A naïve program in MyBase for computing $\sqrt{x}$

```
101 program {
102  procedure SQRT(radicand) {
103   call SqrtIter(0,radicand,call Precision(radicand
           ))
104  }
105  procedure SqrtIter(approximation,radicand,
         precision) {
106   let
107    bid = call Improve(approximation,radicand,
           precision)
108   in
109    if call IsPreciseEnough?(bid,radicand)
110    then bid
111    else call SqrtIter(bid,radicand,precision)
112  }
113  procedure Improve(approximation,radicand,
         precision) {
114   +(approximation,precision)
115  }
116  procedure Precision(x) {1}
117  procedure IsPreciseEnough?(root,square) {
118   lt?(square,call Square(root))
119  }
120  procedure Square(x) {*(x,x)}
121  procedure Abs(x) {if lt?(x,0) then -(0,x) else x}
122  procedure main() {call SQRT(5)}
123 }
```

Listing 7: Advice in HisExt$_1$ for using Newton's method

```
201 aop1 {
202  around: and(pexecution(Improve) args(an,x,epsilon)) {
203   /(+(an,/(x,an)),2)
204  }
205  around: and(pexecution(IsPreciseEnough?) args(root,x)
         ) {
206   lt? (call Abs(-(x,call Square(root))),call
           Precision(x))
207  }
208  around : pcall(Precision) {
209   /(proceed,1000)
210  }
211 }
```

Listing 8: Advice in HerExt$_2$ for preventing an exception

```
301 aop2 { guard_cflow SQRT resume_with call Precision(1) }
```

Newton's method:
$$a_{n+1} = \frac{1}{2}\left(a_n + \frac{x}{a_n}\right)$$

The second around advice (lines 205–207) intercepts `IsPreciseEnough?` executions and checks instead whether or not $\left|(a_n)^2 - x\right| < \epsilon$ where $\epsilon = \frac{1}{1000}$ is set in the third around advice (lines 208–211). The successive approximations now converge quadratically.

Running `main` and calling

```
call SQRT(5)
```

returns [7]

---

[7]The result shown is the actual value returned by the Scheme [36] implementation.

```
(num-val 161/72)
```

meaning $\frac{161}{72} = 2.2361111 = \sqrt{5.0001929} \doteq \sqrt{5}$.

The improved program works well for all non-negative inputs to `SQRT`, except for when the radicand is 0. In this case, `Improve` is called with the first argument $a_n$ set to 0. The execution of `Improve` triggers the advice around `Improve` execution which divides $x$ by $a_n$. Since the value of $a_n$ is 0 an exception occurs.

### 3.3 Third-party Composition

The main point of this example is that HisExt$_1$ and HerExt$_2$ are subject to third-part composition with MyBase and work collaboratively:

- **Units of independent production.** HisExt$_1$ and HerExt$_2$ are independently constructed.

- **Units of composition.** MyBase, HisExt$_1$, and HerExt$_2$ are units of composition. MyBase can be used by itself (running only Listing 6). MyBase can be used with HisExt$_1$ alone (omitting Listing 8). MyBase can be used with HerExt$_2$ alone (omitting Listing 7). MyBase can be used with both HisExt$_1$ and HerExt$_2$.

- **Units of collaboration.** When HisExt$_1$ and HerExt$_2$ are both used they collaborate. In the absence of HerExt$_2$, calling

  ```
  call SQRT(0)
  ```

  results in

  ```
  Error in /: undefined for 0.
  ```

  However, when HerExt$_2$ with the advice code in Listing 8 are present, the correct value 0 is returned. The violating primitive division application is introduced by the advice of HisExt$_1$, yet intercepted by the advice of HerExt$_2$. This desired behavior is non-trivial because HisExt$_1$ was constructed without any prior knowledge of HerExt$_2$.

### 3.4 Analysis

In order to achieve a correct collaboration:

- The aspectual effect of all extension programs needs to be exposed to all the collaborating aspect mechanisms.

- Each individual aspect mechanism must hide its implementation from other aspect mechanisms.

#### 3.4.1 *Exposure of Aspectual Effect*

In the context of multiple distinct aspect mechanisms, certain elements of the aspect program should be exposed to all collaborating aspect mechanisms. We call these elements the *aspectual effect*. The aspectual effect of an aspect program generally specifies the implementation of a crosscutting concern. We assume that the aspectual effect is expressed in the base language.

In our example, the aspectual effect of an $aspect_1 \in$ HisExt$_1$ is specified by advice-body expressions; the aspectual effect of an $aspect_2 \in$ HerExt$_2$ is specified by handler expressions. When HisExt$_1$ and HerExt$_2$ are composed together, their mechanisms must reflect each other's effect. Specifically, HisExt$_1$ aspects must be able to advise procedure calls made from the HerExt$_2$ handler expressions; and HerExt$_2$ handlers must be able to intercept exceptions introduced by the HisExt$_1$ pieces of advice.

### 3.4.2 Hiding of Mechanism Implementation

An aspect extension extends the base language with new functionality. For example, HisExt$_1$ adds advice binding, and HerExt$_2$ adds exception handling to the base language. An aspect mechanism that implements the new functionality must hide its internal operations from the other aspect mechanisms. In our example, pointcut matching and advice selection operations of the HisExt$_1$ mechanism must be hidden from the HerExt$_2$ mechanism. Conversely, testing whether the second division primitive argument evaluates to zero and the exception handler selection of HerExt$_2$ should be invisible to the HisExt$_1$ mechanism.

## 4. OUR APPROACH

Now that we have illustrated a desired behavior, we explain our solution to the aspect mechanism composition problem in general.

### 4.1 Aspect Mechanisms as Mixins

The primary idea is to view an aspect mechanism that extends a base mechanism as a *mixin* [13] that is applied to the base mechanism description. A description of a mechanism is an encoding of its implementation (e.g., a configuration of an abstract machine or its semantics). An *aspect mixin mechanism* transforms some of the base mechanism description and introduces some additional description.[8]

By keeping a clean separation between the descriptions of the base and aspect mechanisms, the aspect mixin mechanism may be composed with other mechanisms that extend the same base language. The particular composition strategy may differ. In the next section we show a concete instance of this general approach.

### 4.2 Solution Instance

We illustrate the approach specifically for expression evaluation semantics. To build a multi mechanism, the composed aspect mechanisms are organized in a chain-of-responsibility [20], pipe-and-filter architecture [37] (Figure 4). Each aspect mechanism performs some part of the evaluation and forwards other parts of the evaluation to the next mechanism using delegation semantics [8] ("super"-like calls). If an expression is delegated by all mechanisms then it is eventually evaluated in $\mathcal{B}$. All the mechanisms defer to $\mathcal{A}$ for the evaluation of recursive and other "self"-calls.

A subtlety in designing a collaborative aspect mechanism is deciding what to hide, what to delegate, and what to expose. A mechanism may hide its effect by directly reducing an expression. A mechanism may refine the next mechanism's semantics by delegating the evaluation. A mechanism may expose its effect by evaluating expressions in $\mathcal{A}$. The latter allows what is known as "weaving". The exposed expressions are then evaluated collaboratively by all the mechanisms. As a result, an effect of an aspect mechanism is made visible to all the other mechanisms. Hence, the mechanisms reflect one another's effect. Overall, a mechanism is con-sidered a collaborative unit provided it properly hides, delegates, or exposes the evaluation.

---
[8]We generally assume that granularity requirements of an aspect mechanism can always be satisfied by either taking the most fine-grained description form (e.g., small-step operational semantics), or refining (e.g., annotating) the current description.

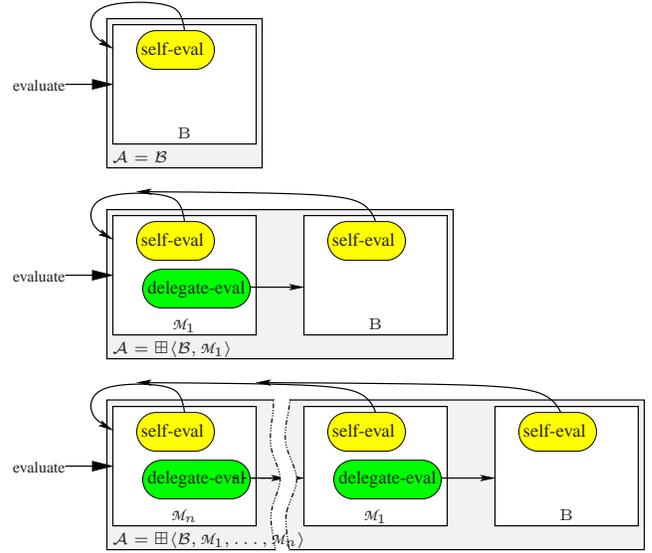

Figure 4: Mixing-like composition of aspect mechanisms

*Notation.*

The following notations are pertinent. We express functions in Curried form. The Curried function definition

$$\textbf{fn}\ pat_1\ pat_2 \ldots pat_n \Rightarrow exp$$

is the same as the lambda expression $\lambda\ pat_1.\lambda\ pat_2.\ldots.\lambda\ pat_n.exp$. Correspondingly, we write a list of function arguments with no parentheses or commas to express a function application that takes the first argument as its single parameter, which could be a tuple, constructs and returns a new function, which then takes the next argument as its single parameter, and so on. In function types, '$\to$' associates to the right.

We use the form $(id\ \textbf{as}\ pat)$ in a formal argument to bind an identifier $id$ to a value and match the value with a pattern $pat$. Variables in the pattern bind to their corresponding values. We use $\textbf{val}\ pat = val$ to split apart a value. The symbol '\_' stands for an anonymous variable (don't care). The symbol $\diamond$ denotes an empty mapping and $[]$ denotes an empty list.

### 4.2.1 Overall Semantics

Let $\mathcal{A}[\![exp]\!]$ denote the meaning of an AOP expression $exp$. Our goal is to be able to build the multi mechanism $\mathcal{A}$ by composing the base mechanism $\mathcal{B}$ and the mutually independent aspect mechanisms $\mathcal{M}_1, \ldots, \mathcal{M}_n$.

Base introduces a domain $\textbf{Exp}_0$ of base expressions. In addition, each of the extensions $\text{Ext}_1, \text{Ext}_2, \ldots, \text{Ext}_n$ may introduce its own respective domain of additional expressions $\textbf{Exp}_1, \textbf{Exp}_2, \ldots, \textbf{Exp}_n$.[9] The domain of AOP expressions $\textbf{Exp}_A$ is hence a union of pairwise disjoint expression domains defined by:

$$\textbf{Exp}_A = \textbf{Exp}_0 + \textbf{Exp}_1 + \textbf{Exp}_2 + \cdots + \textbf{Exp}_n$$

---
[9]We assume that $\textbf{Exp}_i \cap \textbf{Exp}_j = \phi$ for all $0 \le i < j \le n$.

The additional expressions are concern *integration* instructions for the respective aspect mechanism. A concern *implementation*, on the other hand, is expressed using base language expressions in $\mathbf{Exp}_0$ only.

EXAMPLE 1. *$\mathsf{HisExt}_1$ introduces a proceed-exp and $\mathsf{HerExt}_2$ a raise-exp to specify nesting of advice and handler executions, respectively. An $aspect_1 \in \mathsf{HisExt}_1$ in implemented in $\mathbf{Exp}_0 + \{proceed\text{-}exp\}$ and an $aspect_2 \in \mathsf{HerExt}_2$ in $\mathbf{Exp}_0 + \{raise\text{-}exp\}$.*

We use the term *AOP configuration* to denote the state of a multi mechanism $\mathcal{A}$. An AOP configuration $cfg \in \mathbf{Cfg}_A$ is a vector of configurations of the composed mechanisms:

$$\mathbf{Cfg}_A = \mathbf{Cfg}_0 \times \mathbf{Cfg}_1 \times \mathbf{Cfg}_2 \times \cdots \times \mathbf{Cfg}_n$$

where $\mathbf{Cfg}_0$ denotes a domain of the base mechanism states, and $\mathbf{Cfg}_i, 1 \leq i \leq n$, denotes a domain of the aspect mechanism $\mathcal{M}_i$ states.

EXAMPLE 2. *Informally, a $\mathsf{MyBase}$ mechanism configuration comprises a procedure environment, a variable environment, and a store. A $\mathsf{HisExt}_1$ mechanism configuration comprises a list of advice, a "current" join point, and a "current" proceed computation.*

The effect of evaluating an expression $exp \in \mathbf{Exp}_A$ is to change the AOP configuration. The meaning of an expression $exp \in \mathbf{Exp}_A$, denoted $\mathcal{A}[\![exp]\!]$, is defined to be a partial function on configurations:

$$\mathcal{A} : \mathbf{Exp}_A \to \overbrace{(\mathbf{Cfg}_A \hookrightarrow \mathbf{Cfg}_A)}^{\mathbf{Cont}_A}$$

We denote by $\mathbf{Cont}_A$ the set of partial functions on $\mathbf{Cfg}_A$.

### 4.2.2 Design Guidelines for the Base Mechanism

$\mathcal{B}$ provides semantics for expressions in $\mathsf{Base}$. The meaning of an expression $exp \in \mathbf{Exp}_0$ in $\mathsf{Base}$, denoted $\mathcal{B}[\![exp]\!]$, is expected to be defined as:

$$\mathcal{B} : \mathbf{Exp}_0 \to \mathbf{Cont}_A$$

The semantical function $\mathcal{B}$ should adhere to the following design principles:

- All sub-reductions within a $\mathcal{B}$-reduction are reduced by calling $\mathcal{A}$ instead of $\mathcal{B}$.
- $\mathcal{B}$ only accesses and updates the head $\mathbf{Cfg}_0$-element of the $cfg \in \mathbf{Cfg}_A$ configuration, and carries the tail through the computation.

Note that the fact that $\mathcal{B}$ is defined in terms of $\mathbf{Cfg}_A$ does not mean that $\mathcal{A}$ or $n$ are known at the time of writing $\mathcal{B}$. At the time of writing the base mechanism, $\mathcal{A}$ is assumed to be:

$$\mathcal{A}[\![exp]\!] = \begin{cases} \mathcal{B}[\![exp]\!] & exp \in \mathbf{Exp}_0 \\ \bot & \text{otherwise} \end{cases}$$

where $\bot$ stands for "undefined". Let $\hat{\mathcal{B}} : \mathbf{Exp}_0 \to \mathbf{Cfg}_0 \to \mathbf{Cfg}_0$ denote the evaluation semantics for $\mathsf{Base}$ with its standard signature. $\hat{\mathcal{B}}$ is extended to have the signature of $\mathcal{B}$ (without knowing $n$) as follows: $\forall exp \in \mathbf{Exp}_0, \forall cfg = cfg_0 :: cfg^* \in \mathbf{Cfg}_A$:

$$\mathcal{B}[\![exp]\!] \, cfg = \begin{cases} cfg'_0 :: cfg^* & \hat{\mathcal{B}}[\![exp]\!] \, cfg_0 = cfg'_0 \\ \bot & \hat{\mathcal{B}}[\![exp]\!] \, cfg_0 = \bot \end{cases}$$

### 4.2.3 Design Guidelines for an Aspect Mechanism

We construct the aspect mechanism $\mathcal{M}_i$ for an aspect extension $\mathsf{Ext}_i$ as the override combination[10] of a semantics transformer $\mathcal{T}_i$ and a semantical function $\mathcal{E}_i$:

$$\mathbf{val} \; \mathcal{M}_i = \mathbf{fn} \; eval \Rightarrow (\mathcal{T}_i \; eval) \oplus \mathcal{E}_i$$

Semantics for the $\mathsf{Ext}_i$'s newly introduced expression domain $\mathbf{Exp}_i$ is defined by:

$$\mathcal{E}_i : \mathbf{Exp}_i \to \mathbf{Cont}_A$$

The introduction of $\mathsf{Ext}_i$ into the base language also requires a change to the evaluation semantics for a non-empty[11] subset of the existing base language expressions $\mathbf{Exp}_0^i \subseteq \mathbf{Exp}_0$. We define this part of the semantics for $\mathsf{Ext}_i$ as a language semantics transformer:

$$\mathcal{T}_i : \overbrace{(\mathbf{Exp}_0 \to \mathbf{Cont}_A)}^{\mathbf{Eval}_0} \to \overbrace{(\mathbf{Exp}_0^i \to \mathbf{Cont}_A)}^{\mathbf{Eval}_0^i}$$

The semantics transformer $\mathcal{T}_i$ should adhere to the following design principles:

- $\mathcal{T}_i$ defines the semantics for $\mathsf{Ext}_i$ and nothing more. Let $\mathcal{B}'$ denote a semantical function with the same signature as $\mathcal{B}$ or an extended signature.[12] $\mathcal{T}_i(\mathcal{B}')$ delegates the evaluation to $\mathcal{B}'$ whenever the base language semantics is required.
- $\mathcal{T}_i(\mathcal{B}')$ accesses only the $\mathbf{Cfg}_0$- and $\mathbf{Cfg}_i$-elements in a $cfg \in \mathbf{Cfg}_A$ configuration, while the rest are carried through the computation.

Note that allowing the aspect mechanism access to the $\mathbf{Cfg}_0$ element is needed for modeling interesting cases of aspect mechanism interactions.

### 4.2.4 Third-party Construction of an AOP Language

Let $\mathcal{B}$ denote the $\mathsf{Base}$ mechanism, and let $\{k_i\}_{i=1}^n$ be an ordered index set. Let $\mathcal{M}_{k_1}, \ldots, \mathcal{M}_{k_n}$ denote the aspect mechanisms for the aspect extensions $\mathsf{Ext}_{k_1}, \ldots, \mathsf{Ext}_{k_n}$, respectively.

We construct the multi mechanism $\mathcal{A}$ as the composition:

$$\mathcal{A} = \boxplus \langle \mathcal{B}, \mathcal{M}_{k_1}, \ldots, \mathcal{M}_{k_n} \rangle$$

where the composition semantics for $\boxplus$ is defined as following. The meaning of $exp \in \mathbf{Exp}_A$, denoted $\mathcal{A}_n[\![exp]\!] \, cfg$, is given by the recurrence relation:

$$\mathcal{A}_0 = \mathcal{B}$$
$$\mathcal{A}_n = \mathcal{A}_{n-1} \oplus (\mathcal{M}_{k_n} \; \mathcal{A}_{n-1})$$

By construction,

$$\mathcal{A}_n : (\mathbf{Exp}_0 + \mathbf{Exp}_{k_1} + \cdots + \mathbf{Exp}_{k_n}) \to \mathbf{Cont}_A$$

is of the right signature and obeys the composition principle. To illustrate the construction, we conclude by elaborating the first three instances:

---

[10]For two partial functions $g$ and $h$, their override combination $g \oplus h$ ($h$ overrides $g$), is defined by:

$$(g \oplus h)(x) =_{def} \begin{cases} h(x) & x \in \mathbf{dom} \; h \\ g(x) & otherwise \end{cases}$$

[11]W.l.o.g., assume $\mathbf{Exp}_0^i \neq \phi$.

[12]An extended $\mathcal{B}$ may have a signature $\mathcal{B}' : \mathbf{Exp}_0' \to \mathbf{Cfg}_A$, where $\mathbf{Exp}_0' \supseteq \mathbf{Exp}_0$.

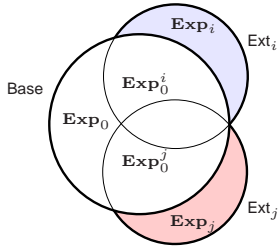

Figure 5: Expression Domains for $l = 2$

- For $l = 0$, we have that $\mathbf{Exp}_A = \mathbf{Exp}_0$, and the meaning of $exp \in \mathbf{Exp}_A$ is the same as the meaning of $exp$ in Base:

$$\mathcal{A}^\phi : \mathbf{Exp}_0 \to \mathbf{Cont}_A$$

$$\mathcal{A}^\phi[\![exp]\!]\, cfg = \mathcal{B}[\![exp]\!]\, cfg$$

- For $l = 1$ and the singleton index set $\{i\}$ for some $1 \leq i \leq n$, we have that $\mathbf{Exp}_A = \mathbf{Exp}_0 + \mathbf{Exp}_i$. The meaning of $exp \in \mathbf{Exp}_A$ is

$$\mathcal{A}^{\{i\}} : (\mathbf{Exp}_0 + \mathbf{Exp}_i) \to \mathbf{Cont}_A$$

We construct:

$$\mathcal{A}^{\{i\}} = \mathcal{B} \oplus \overbrace{(\mathcal{T}_i\, \mathcal{B})}^{\mathcal{M}_i\, \mathcal{B}} \oplus \mathcal{E}_i$$

$$\mathcal{A}^{\{i\}}[\![exp]\!]\, cfg = \begin{cases} \mathcal{E}_i[\![exp]\!]\, cfg & exp \in \mathbf{Exp}_i \\ (\mathcal{T}_i\, \mathcal{B})[\![exp]\!]\, cfg & exp \in \mathbf{Exp}_0^i \\ \mathcal{B}[\![exp]\!]\, cfg & otherwise \end{cases}$$

- For $l = 2$ and the ordered index set $\{i, j\}$ for some $1 \leq i, j \leq n$, we have that $\mathbf{Exp}_A = \mathbf{Exp}_0 + \mathbf{Exp}_i + \mathbf{Exp}_j$ (Figure 5). The meaning of $exp \in \mathbf{Exp}_A$ is

$$\mathcal{A}^{\{i,j\}} : (\mathbf{Exp}_0 + \mathbf{Exp}_i + \mathbf{Exp}_j) \to \mathbf{Cont}_A$$

We construct:

$$\mathcal{A}^{\{i,j\}} = \mathcal{A}^{\{i\}} \oplus \overbrace{(\mathcal{T}_j\, \mathcal{A}^{\{i\}})}^{\mathcal{M}_j\, \mathcal{A}^{\{i\}}} \oplus \mathcal{E}_j$$

$$\mathcal{A}^{\{i,j\}}[\![exp]\!] = \begin{cases} \mathcal{E}_j[\![exp]\!] & exp \in \mathbf{Exp}_j \\ \mathcal{E}_i[\![exp]\!] & exp \in \mathbf{Exp}_i \\ (\mathcal{T}_j\, \mathcal{B})[\![exp]\!] & exp \in \mathbf{Exp}_0^j - \mathbf{Exp}_0^i \\ (\mathcal{T}_i\, \mathcal{B})[\![exp]\!] & exp \in \mathbf{Exp}_0^i - \mathbf{Exp}_0^j \\ (\mathcal{T}_j\, (\mathcal{B} \oplus (\mathcal{T}_i\, \mathcal{B})))[\![exp]\!] & exp \in \mathbf{Exp}_0^i \cap \mathbf{Exp}_0^j \\ \mathcal{B}[\![exp]\!] & otherwise \end{cases}$$

## 5. IMPLEMENTATION

As a proof of concept we have implemented MyBase, HisExt$_1$, and HerExt$_2$ for the example presented in Section 3. This section provides the implementation details more formally to the so-inclined reader.

### 5.1 Base Mechanism Implementation

The domain $\mathbf{Exp}_A$ of AOP expressions includes MyBase, HisExt$_1$, and HerExt$_2$ expressions. We define $\mathbf{Exp}_0$ by extending the expression set $\mathbf{Exps}$ with a set of annotated expressions:

$$\mathbf{Exp}_0 = \mathbf{Exps} + annotated\text{-}exp$$

| annotated-exp | = | procbody-exp \|procarg-exp \|primarg-exp \| |
| | | assignrhs-exp \|block-exp \|letbody-exp \| |
| | | letrhs-exp \|if-exp \|then-exp \|else-exp |
| procbody-exp | = | $\mathbf{Exp}_0 \times \mathbf{PNm}$ | Procedure body |
| procarg-exp | = | $\mathbf{Exp}_0 \times (\mathbf{PNm} \times \mathbf{Var})$ | Procedure arg |
| primarg-exp | = | $\mathbf{Exp}_0 \times (Prim \times Int)$ | Primitive arg |
| assignrhs-exp | = | $\mathbf{Exp}_0 \times \mathbf{Var}$ | Assignment RHS |
| block-exp | = | $\mathbf{Exp}_0 \times Int$ | Block element |
| letbody-exp | = | $\mathbf{Exp}_0 \times \mathbf{Var}^*$ | Let body |
| letrhs-exp | = | $\mathbf{Exp}_0 \times (\mathbf{Var} \times Int)$ | Let env RHS |
| if-exp | = | $\mathbf{Exp}_0 \times \{\texttt{if}\}$ | If exp |
| then-exp | = | $\mathbf{Exp}_0 \times \{\texttt{then}\}$ | Then exp |
| else-exp | = | $\mathbf{Exp}_0 \times \{\texttt{else}\}$ | Else exp |

Figure 6: Annotated Expressions

| app-exp | = | $\mathbf{PNm} \times procarg\text{-}exp^*$ | Procedure call |
| begin-exp | = | $block\text{-}exp^*$ | Block |
| cond-exp | = | $if\text{-}exp \times then\text{-}exp \times else\text{-}exp$ | Conditional exp |
| assign-exp | = | $\mathbf{Var} \times assignrhs\text{-}exp$ | Assignment |
| let-exp | = | $\mathbf{Var}^* \times letrhs\text{-}exp^* \times$ | |
| | | $letbody\text{-}exp$ | Let |
| primapp-exp | = | $Prim \times primarg\text{-}exp^*$ | Primitive app |

Figure 7: Complex Expressions

| $cfg_0 \in \mathbf{Cfg}_0$ | = | $\mathbf{Env}_P \times \mathbf{Env}_V \times$ | Base |
| | | $\mathbf{Store}$ | configuration |
| $env_V \in \mathbf{Env}_V$ | = | $\mathbf{Var} \to \mathbf{Loc}$ | Variable envs |
| $sto \in \mathbf{Store}$ | = | $\mathbf{Loc} \to \mathbf{Val}$ | Value Stores |
| $env_P \in \mathbf{Env}_P$ | = | $\mathbf{PNm} \to \mathbf{Proc}$ | Procedure envs |
| $\theta \in \mathbf{Proc}$ | = | $\mathbf{Var}^* \times procbody\text{-}exp$ | Procedures |

Figure 8: MyBase domains

Annotated expressions (Figure 6) extend the interface of the base mechanism to satisfy granularity needs of the HisExt$_1$ and HerExt$_2$ mechanisms. A complex expression (Figure 7) includes annotated expressions as subexpressions.

The base configuration domain $\mathbf{Cfg}_0$ consist of a procedure environment domain $\mathbf{Env}_P$, a variable environment domain $\mathbf{Env}_V$, and a value store domain $\mathbf{Store}$ (Figure 8). A procedure is represented as a closure that contains argument names and a procedure body expression. The other definitions are omitted.

The evaluation semantics $\mathcal{B}$ (Figure 9) for $\mathbf{Exp}_0$ expressions satisfies the design principles for the base mechanisms: (1) all expression evaluations in $\mathcal{B}$ are *exposed* to $\mathcal{A}$ (highlighted in the figure); (2) it accesses and updates only the $\mathbf{Cfg}_0$-element of the configuration; (3) the other configurations are carried through the computation.

### 5.2 Aspect Mechanism Implementation

The aspect mechanisms are implemented as mixins to the base mechanism (Figure 10). The semantics for Ext$_i$ is specified using three constructor functions:

- *build-*$\mathcal{E}_i$ constructs an evaluator for $\mathbf{Exp}_i$ expressions:

$$\textit{build-}\mathcal{E}_i : Int \to (\mathbf{Exp}_i \to \mathbf{Cont}_A)$$

$$\begin{aligned}
&\textbf{val } \mathcal{B} : \textbf{Exp}_0 \to \textbf{Cont}_A \\
&= \textbf{fn } (\texttt{lit-exp} \langle num \rangle) \langle \_, \_, sto \rangle :: cfg^* \Rightarrow \\
&\quad \langle \diamond, \diamond, sto[0 \mapsto (num\text{-}val\ num)] \rangle :: cfg^* \\
&\mid \textbf{fn } (\texttt{true-exp} \langle \rangle) \langle \_, \_, sto \rangle :: cfg^* \Rightarrow \\
&\quad \langle \diamond, \diamond, sto[0 \mapsto (bool\text{-}val\ \#\texttt{t})] \rangle :: cfg^* \\
&\mid \textbf{fn } (\texttt{false-exp} \langle \rangle) \langle \_, \_, sto \rangle :: cfg^* \Rightarrow \\
&\quad \langle \diamond, \diamond, sto[0 \mapsto (bool\text{-}val\ \#\texttt{f})] \rangle :: cfg^* \\
&\mid \textbf{fn } (\texttt{app-exp} \langle pname, [exp_1, \ldots, exp_n] \rangle)\ cfg_0 :: cfg^* \\
&\quad \textbf{let} \\
&\quad\quad \textbf{val } \langle env_P, env_V, sto \rangle = cfg_0 \\
&\quad\quad \textbf{val } \langle \_, \_, sto_1 \rangle :: cfg_1^* = \\
&\quad\quad\quad \mathcal{A}\ exp_1\ \langle env_P, env_V, sto \rangle :: cfg^* \\
&\quad\quad \ldots \\
&\quad\quad \textbf{val } \langle \_, \_, sto_n \rangle :: cfg_n^* = \\
&\quad\quad\quad \mathcal{A}\ exp_n\ \langle env_P, env_V, sto_{n-1} \rangle :: cfg_{n-1}^* \\
&\quad\quad \textbf{val } \langle [id_1, \ldots, id_n], exp_{proc} \rangle = env_P\ pname \\
&\quad\quad \textbf{val } v_1 = sto_1\ 0 \\
&\quad\quad \ldots \\
&\quad\quad \textbf{val } v_n = sto_n\ 0 \\
&\quad\quad \textbf{val } sto_{n+1} = sto_n[l_1 \mapsto v_1],\ l_1 \notin \textbf{dom } sto_n \\
&\quad\quad \ldots \\
&\quad\quad \textbf{val } sto_{2n} = sto_{2n-1}[l_n \mapsto v_n],\ l_n \notin \textbf{dom } sto_{2n-1} \\
&\quad \textbf{in} \\
&\quad\quad \mathcal{A}\ exp_{proc}\ \langle env_P, \diamond[id_1 \mapsto l_1, \ldots, id_n \mapsto l_n], sto_{2n} \rangle :: cfg_n^* \\
&\quad \textbf{end} \\
&\mid \ldots \\
&\mid \textbf{fn } (\texttt{annotated-exp} \langle exp, \_ \rangle)\ cfg \Rightarrow \mathcal{A}\ exp\ cfg
\end{aligned}$$

Figure 9: MyBase semantical function

- *build-$\mathcal{T}_i$* constructs the semantics transformer for the $\mathsf{Ext}_i$:

$$build\text{-}\mathcal{T}_i : Int \to \overbrace{(\textbf{Exp}_0 \to \textbf{Cont}_A) \to (\textbf{Exp}_0^i \to \textbf{Cont}_A)}^{\textbf{Eval}_0 \to \textbf{Eval}_0^i}$$

- *build-$\mathcal{M}_i$* constructs the aspect mixin mechanism $\mathcal{M}_i$ for $\mathsf{Ext}_i$:

$$\begin{aligned}
&\textbf{val } build\text{-}\mathcal{M}_i : Int \to \textbf{Eval}_0 \to (\textbf{Exp}_0^i + \textbf{Exp}_i) \to \textbf{Cont}_A \\
&= \textbf{fn } pos\ eval \Rightarrow (build\text{-}\mathcal{T}_i\ pos\ eval) \oplus (build\text{-}\mathcal{E}_i\ pos)
\end{aligned}$$

The $Int$ arguments provides the position of the extension's configuration domain $\textbf{Cfg}_i$ within $\textbf{Cfg}_A$.

### 5.2.1 $\mathsf{HisExt}_1$ *Mechanism*

The aspect mechanism $\mathcal{M}_1$ transforms the semantics for procedure calls and executions, and supplies semantics for $\textbf{Exp}_1$'s new proceed expression:

$$\begin{aligned}
\textbf{Exp}_0^1 &= \{app\text{-}exp, procbody\text{-}exp\} \\
\textbf{Exp}_1 &= \{proceed\text{-}exp\}
\end{aligned}$$

A configuration $cfg_1 \in \textbf{Cfg}_1$ for $\mathsf{HisExt}_1$ (Figure 11) comprises a set of advice, a "current" join point, and a "current" proceed continuation. An advice $adv \in \textbf{Adv}$ is derived directly from $\mathsf{HisExt}_1$'s syntax. A join point $jp \in \textbf{JP}$ is an abstraction of the procedure call stack. It stores the name, formal and actual arguments of a corresponding procedure. The third element provides a meaning for proceed expressions. The effect and binding domains are internal to the mechanism. An effect carries a set of bindings and an advice

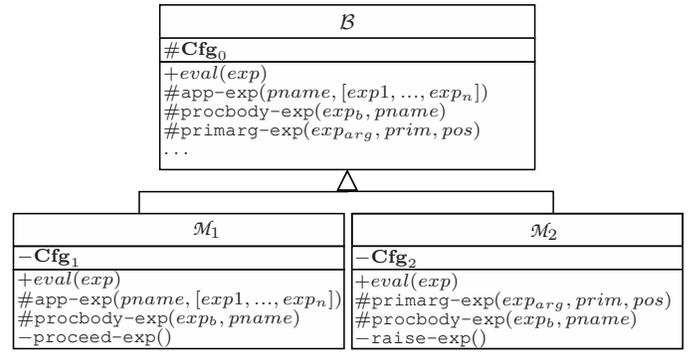

Figure 10: Aspect mechanisms as mixins

$$\begin{array}{lll}
exp \in \textbf{Exp}_{adv} & = \textbf{Exp}_0 + \textbf{Exp}_1 & \text{Advice exps} \\
cfg_1 \in \textbf{Cfg}_1 & = \textbf{Adv}^* \times \textbf{JP} \times \textbf{Cont}_A & \text{Configuration} \\
adv \in \textbf{Adv} & = \textbf{PCD} \times \textbf{Exp}_{adv} & \text{Advice} \\
jp \in \textbf{JP} & = \{\texttt{call}, \texttt{exec}\} \times \textbf{PNm} \times & \\
& \quad \textbf{Var}^* \times \textbf{Val}^* \times \textbf{JP} + \textbf{Unit} & \text{Join points} \\
pcd \in \textbf{PCD} & & \text{Pointcuts} \\
effect \in \textbf{Effect} & = \textbf{Bnd}^* \times \textbf{Exp}_{adv} & \text{Effects} \\
bnd \in \textbf{Bnd} & = \textbf{Var} \times \textbf{Val} & \text{Binding}
\end{array}$$

Figure 11: $\mathsf{HisExt}_1$ Domains

$$\begin{aligned}
&\textbf{local} \\
&\textbf{val } app\text{-}eff : Int \to \textbf{Effect}^* \to \textbf{Eval}_0 \to \textbf{Eval}_0 \\
&= \textbf{fn } \_\ []\ eval\ exp\ cfg \Rightarrow eval\ exp\ cfg \\
&\mid \textbf{fn } i\ \langle bnd_{adv}^*, exp_{adv} \rangle :: effect^*\ eval \Rightarrow \\
&\quad \textbf{fn } exp\ \langle env_P, env_V, sto \rangle :: cfg^* \Rightarrow \\
&\quad \textbf{let} \\
&\quad\quad \textbf{val } \langle adv^*, jp, procd \rangle = \pi_i(cfg^*) \\
&\quad\quad \textbf{val } procd' : \textbf{Cont}_A \\
&\quad\quad = \textbf{fn } \langle \_, \_, sto' \rangle :: cfg^{*'} \Rightarrow \\
&\quad\quad\quad app\text{-}eff\ i\ effect^*\ eval\ exp\ \langle env_P, env_V, sto' \rangle :: cfg^{*'} \\
&\quad\quad \textbf{val } \langle env_V{}', sto' \rangle = build\text{-}adv\text{-}env\ bnd_{adv}^*\ sto \\
&\quad\quad \textbf{val } cfg^{*'} = cfg^*[i \mapsto \langle adv^*, jp, procd' \rangle] \\
&\quad\quad \textbf{val } cfg_0' :: cfg^{*''} = \mathcal{A}\ exp_{adv}\ \langle env_P, env_V{}', sto' \rangle :: cfg^{*'} \\
&\quad \textbf{in} \\
&\quad\quad cfg_0' :: cfg^{*''}[i \mapsto \langle adv^*, jp, procd \rangle] \\
&\quad \textbf{end} \\
&\ldots \\
&\textbf{in} \\
&\textbf{val } build\text{-}\mathcal{T}_1 : Int \to \textbf{Eval}_0 \to \textbf{Eval}_0^1 \\
&= \textbf{fn } i\ eval\ exp\ cfg_0 :: cfg^* \Rightarrow \\
&\textbf{let} \\
&\quad \textbf{val } \langle adv^*, jp_{enc}, procd \rangle = \pi_i(cfg^*) \\
&\quad \textbf{val } jp = build\text{-}jp\ exp\ jp_{enc}\ cfg_0 \\
&\quad \textbf{val } effect^* = match\text{-}jp\ jp\ adv^* \\
&\quad \textbf{val } cfg^{*'} = cfg^*[i \mapsto \langle adv^*, jp, procd \rangle] \\
&\quad \textbf{val } cfg_0' :: cfg^{*''} = app\text{-}eff\ i\ effect^*\ eval\ exp\ cfg_0 :: cfg^{*'} \\
&\textbf{in} \\
&\quad cfg_0' :: cfg^{*''}[i \mapsto \langle adv^*, jp_{enc}, procd \rangle] \\
&\textbf{end} \\
&\textbf{end}
\end{aligned}$$

Figure 12: *build-$\mathcal{T}_1$* semantics

```
val build-𝓔₁ : Int → Exp₁ → Cont_A
 = fn i (proceed-exp ⟨⟩) (cfg as _ :: cfg*) ⇒
   let
     val ⟨_, _, procd⟩ = π_i(cfg*)
   in
     procd cfg
   end
```

Figure 13: *build-𝓔₁* semantics

body expression. The bindings provide an appropriate variable environment for evaluating the advice body expression.

The interesting part of the aspect mechanism $\mathcal{M}_1$ implementation is given by *build-$\mathcal{T}_1$* (Figure 12). *build-$\mathcal{T}_1$* defines a transformer of the semantics for procedure calls and procedure executions. The new semantics creates a join point, matches it against an advice list, and applies selected advice effects in *app-eff*. The function ensures that the mechanism's configuration properly reflects a "current" join point by setting it before and after an effect application.

*app-eff* has two general behaviors. If the effect list is empty then the expression evaluation is *delegated*. Otherwise, the function *exposes* the effect by evaluating the advice expression $exp_{adv}$ in $\mathcal{A}$. $exp_{adv}$ is evaluated in a properly constructed variable environment $env_{V\,adv}$ and a proceed continuation $procd'$.

*app-eff* ensures that the mechanism configuration always stores a proper proceed continuation in the same manner as *build-$\mathcal{T}_1$* reflects a "current" join point. This makes *build-$\mathcal{E}_1$* straightforward (Figure 13). The meaning of a `proceed-exp` expression is given by the proceed continuation obtained from the configuration. The continuation then runs *app-eff* on the rest of the effect list. In other words, a `proceed-exp` expression either evaluates the next advice in $\mathcal{A}$ or delegates the evaluation to *eval* if there is no advice left.

Due to space considerations, we omit the HisExt₁ functions *match-jp*, *build-jp* and *build-adv-env*, which do not affect the mechanism composition semantics.

### 5.2.2 HerExt₂ Mechanism

The $\mathcal{M}_2$ mechanism for HerExt₂ transforms the semantics for a primitive argument and procedure execution expressions, and supplies semantics for **Exp**₂'s new raise expression:

$$\mathbf{Exp}_0^2 = \{\textit{primarg-exp, procbody-exp}\}$$
$$\mathbf{Exp}_2 = \{\textit{raise-exp}\}$$

A configuration $cfg_2 \in \mathbf{Cfg}_2$ (Figure 14) stores a list of handlers, a stack of currently executing procedures (a list of procedure names),

| | | | |
|---|---|---|---|
| $exp \in \mathbf{Exp}_{hnd}$ | = | $\mathbf{Exp}_0 + \mathbf{Exp}_2$ | Handler exps |
| $cfg_2 \in \mathbf{Cfg}_2$ | = | $\mathbf{Handler}^* \times \mathbf{PNm}^* \times$ | |
| | | $\mathbf{Cont}_A$ | Configuration |
| $hnd \in \mathbf{Handler}$ | = | $\mathbf{PNm} \times \mathbf{Exp}_{hnd}$ | Handlers |

Figure 14: HerExt₂ Domains

```
local
  val app-handler : Int → Exp*_{hnd} → Cont_A
   = fn _ [] cfg ⇒ cfg
   | fn i exp :: exp* ⟨env_P, _, sto⟩ :: cfg* ⇒
     let
       val ⟨hnd*, stack, raise⟩ = π_i(cfg*)
       val υ = sto 0
       val raise' : Cont_A
         = fn ⟨env_P, env_V, sto⟩ :: cfg* ⇒
           app-handler exp* ⟨env_P, env_V, sto[0 ↦ υ]⟩ :: cfg*
       val cfg*' = cfg*[i ↦ ⟨hnd*, stack, raise'⟩]
       val cfg'_0 :: cfg*'' = 𝓐 exp ⟨env_P, ⋄, sto⟩ :: cfg*'
     in
       cfg'_0 :: cfg*''[i ↦ ⟨hnd*, stack, raise⟩]
     end
   ...
in
  val build-𝓣₂ : Int → Eval₀ → Eval₀²
   = fn i eval (primarg-exp ⟨exp_{arg}, prim, pos⟩ as exp) cfg ⇒
     let
       val ⟨env_P, env_V, _⟩ :: _ = cfg
       val (cfg' as ⟨_, _, sto⟩ :: cfg*) = eval exp cfg
     in
       if (sto 0 = (num-val 0) ∧ prim = "/" ∧ pos = 2)
       then
         let
           val ⟨hnd*, stack, _⟩ = π_i(cfg*)
           val exp*_{hnd} = match-handler hnd* stack
         in
           app-handler i exp*_{hnd} ⟨env_P, env_V, sto⟩ :: cfg*
         end
       else cfg'
     end
   | fn i eval (procbody-exp ⟨exp_b, pname⟩ as exp) cfg_0 :: cfg* ⇒
     let
       val ⟨hnd*, stack, raise⟩ = π_i(cfg*)
       val cfg*' = cfg*[i ↦ ⟨hnd*, pname :: stack, raise⟩]
       val cfg'_0 :: cfg*'' = eval exp cfg_0 :: cfg*'
     in
       cfg'_0 :: cfg*''[i ↦ ⟨hnd*, stack, raise⟩]
     end
end
```

Figure 15: *build-$\mathcal{T}_2$* semantics

and a "current" raise continuation. A handler $hnd \in \mathbf{Handler}$ is derived from the syntax of HerExt₂. It contains a name of a guarded procedure and a handler expression. A handler expression may contain a `raise-exp` expression.

The new semantics for `primarg-exp` enables the invocation of a handler in an exceptional situation when the second argument of a division primitive evaluates to zero. In this case, *build-$\mathcal{T}_2$* (Figure 15) selects a list of handler expressions using *match-handler* and invokes them using *app-handler*. If no exception occurs, the original semantics is used.

The mechanism reflects the execution stack of its configuration by transforming the semantics for `procbody-exp` expressions. The new semantics simply pushes the stack before and pops it after ap-

```
val build-$\mathcal{E}_2$ : Int → $\mathbf{Exp}_2$ → $\mathbf{Cont}_A$
= fn i (raise-exp ⟨⟩) (cfg as _ :: cfg*) ⇒
let
  val ⟨_, _, raise⟩ = $\pi_i$(cfg*)
in
  raise cfg
end
```

Figure 16: *build-$\mathcal{E}_2$* semantics

plying *eval*.

*app-handler* produces a configuration transformer from a list of handler expressions. If the list is empty then the transformer is the identity function. Otherwise, the configuration is constructed by evaluating in $\mathcal{A}$ the first handler expression. The function also constructs and reflects a raise continuation in the mechanism configuration. The continuation simply applies *app-handler* to the rest of the handlers.

The *build-$\mathcal{E}_2$* function (Figure 16) is similar to *build-$\mathcal{E}_1$*. The meaning of a raise-exp expression is provided by the raise continuation drawn from the configuration.

Due to space considerations, we omit the *match-handler* function of HerExt$_2$. This function bars no affect on the mechanism composition semantics.

## 5.3 Constructing an AOP language

We construct the semantical function for the composed AOP language as follows:

$$\mathcal{A} = \boxplus \langle \mathcal{B}, \mathcal{M}_1, \mathcal{M}_2 \rangle$$

where

$$\mathcal{M}_1 = \textit{build-}\mathcal{M}_1\ 1$$

and

$$\mathcal{M}_2 = \textit{build-}\mathcal{M}_2\ 2$$

The meaning of a program

$$p = \langle \textit{base}, \textit{aspect}_1, \textit{aspect}_2 \rangle$$

in the composed AOP language is defined as:

$$\mathfrak{M}[\![p]\!] = \mathcal{A}\ \textit{exp}_{\textit{main}}\ \langle \textit{cfg}_0, \textit{cfg}_1, \textit{cfg}_2 \rangle$$

such that

$\textit{exp}_{\textit{main}} = (\texttt{app-exp}\ \langle \text{'main}, [] \rangle)$
$\textit{cfg}_0 = \langle \textit{env}_P, \diamond, \diamond \rangle \qquad \textit{env}_P = \mathfrak{D}_0[\![\textit{base}]\!]$
$\textit{cfg}_1 = \langle \textit{adv}^*, \langle\rangle, \diamond \rangle \qquad \textit{adv}^* = \mathfrak{D}_1[\![\textit{aspect}_1]\!]$
$\textit{cfg}_2 = \langle \textit{hnd}^*, [], \diamond \rangle \qquad \textit{hnd}^* = \mathfrak{D}_2[\![\textit{aspect}_2]\!]$

## 6. DISCUSSION AND FUTURE WORK

Our study of constructing an AOP language with multiple aspect extensions opens interesting research questions.

### 6.1 Alternative Collaboration Semantics

The co-existence of multiple aspect extensions raise a question concerning the desired policy of collaboration. The presented solution instance defines the combinator $\boxplus$ operations to "wrap" aspect mechanisms around each other and around the original meaning. This grants the aspect mechanism with complete control over the original meaning and the option to override it. For example, the HisExt$_1$ mechanism might disable the original semantics of *app-exp* and *procbody-exp* expressions when they are advised with no **proceed**. A mechanism can either delegate the expression evaluation to the next mechanism or evaluate the expression itself. In the latter case, the evaluated expression is "filtered" out. We call this a composition with *wrapping* semantics.

Collaboration with wrapping semantics is sensitive to the order of composition. The program example in Listing 9 illustrates a collaboration with wrapping semantics.

Listing 9: Collaboration semantics in $AOP$

```
program {procedure main() { 1 } }
aop1 { around(): pexecution(main) {/(1,0)} }
aop2 { guard_cflow main resume_with 2 }
```

If the AOP language is constructed as

$$\mathcal{A} = \boxplus \langle \mathcal{B}, \mathcal{M}_2, \mathcal{M}_1 \rangle$$

$\mathcal{M}_1$ is applied first and replaces the *procbody-exp* of **main** with the advice body expression. Consequently, $\mathcal{M}_2$ does not observe the execution of **main** in the execution stack and would not guard the division. The program would therefore throws a divide-by-zero exception. On the other hand, if the language is constructed as

$$\mathcal{A} = \boxplus \langle \mathcal{B}, \mathcal{M}_1, \mathcal{M}_2 \rangle$$

the exception is caught.

In wrapping semantics different mechanisms generally reflect different views of the program execution. Alternatively, one can provide a collaboration semantics where all the mechanisms share a unique program view. This can be achieved by decoupling the reification and reflection processes of a mechanism. With such semantics, every expression evaluated in $\mathcal{A}$ is reified by all the mechanisms. The evaluation semantics is then constructed by all the mechanism collaboratively with respect to the ordering. Given this alternative semantics, the program example in Listing 9 would produce no exception independently of the ordering of $\mathcal{M}_1$ and $\mathcal{M}_2$.

### 6.2 Alternative Semantical Operations

We illustrate our approach using expression evaluation semantics. However, the idea of third-party composition of aspect extensions can be realized for other kinds of semantical operations.

Consider a generalized form of a semantical function type:

$$\mathbf{Mean} = \mathbf{OP} \rightarrow \mathbf{Cont}_A$$

where $\mathbf{OP}$ is a domain of operation identifiers. Given $\mathbf{Cfg}_A = \mathbf{Cfg}_0 \times \mathbf{Cfg}^*$, $\mathbf{Mean}$ maps to various operations of MyBase semantics as shown in Table 1. For example, store lookup operation is identified by location. It takes a store and a (dummy) value, and returns a store and a result value. Our approach can be easily redefined to use $\mathbf{Mean}$ instead of expression evaluation semantics.

### 6.3 Other Solution Instances

The specific $\boxplus$ wrapping semantics is only an illustration of our approach in general. In this sections we discuss how alternative solution instances can be constructed.

The wrapping semantics enables to compose arbitrary aspect mechanisms as long as the mechanisms can be defined as mixins to the base mechanism description. However, wrapping does not support complex mechanism compositions. For example, a reasonable composition of AspectJ and AspectWerkz might require that, at each join point, **before** advice in both AspectJ and AspectWerkz aspects are executed before any **around** advice, and finally followed by **after** advice. However, such an AspectJ/AspectWerkz composition is difficult to construct using the wrapping composition semantics.

| Type | OP | $\text{Cfg}_0$ |
|---|---|---|
| **Expr. eval** | Exp | $\text{Env}_P \times \text{Env}_V \times \text{Store}$ |
| **Store upd** | Loc | Store |
| **Store lookup** | Loc | $\text{Store} \times \text{Val}$ |
| **Env upd** | Var | $\text{Env}_V$ |
| **Env lookup** | Var | $\text{Env}_V \times \text{Loc}$ |

Table 1: Semantical operations in MyBase

More complex composition semantics can be provided by imposing additional requirements on the aspect mechanism design. For example, one possible approach is to specify types of aspectual effect that a mechanism can produce. With such a semantics, the overall aspectual effect can be constructed from aspectual effects of the collaborating mechanisms with regard to the effect types.

## 6.4 Other Mechanism Descriptions

Our choice of the mechanism's description style restricts access to the context data. Specifically, a mechanism can only access elements of the current or parent expression, environment, and stores. While this data can be sufficient for implementing the $\text{HisExt}_1$ and $\text{HerExt}_2$ aspect extensions for MyBase, real-world aspect extensions may generally require more information. For example, AspectJ needs access to callee and caller objects to construct a method call join point. Instantiating the approach for a description style that uses an explicit representation of the evaluation context (e.g., using a CEKS machine [15, 16]) would produce a more general solution.

In our solution we used annotated expressions to meet the granularity requirement of $\text{HisExt}_1$ and $\text{HerExt}_2$. The same result can be achieved by using small-step operational semantics for describing the mechanisms. In this case, aspect mechanisms would transform and extend operational semantics rules of the base mechanism.

## 6.5 Application

This work provides a foundation for composing multiple aspect mechanisms. A practical application of this work is to construct an AOP framework that:

1. supports expressiveness that generalizes over concepts and lingual mechanisms of potential source aspect extensions. This requires a generalized aspect mechanism model.

2. meets granularity requirements of any source aspect extension.

3. provides lingual mechanisms for encapsulating constructs that simulate a source aspect mechanism.

4. provides lingual mechanisms for exposing the aspectual effect of the source aspects.

## 7. RELATED WORK
## 7.1 Composing Aspect Extensions

Several authors point out the expressiveness drawback in using a single general-purpose AOP language, and emphasize the usefulness of combining modular domain-specific aspect extensions [14, 22, 48, 38, 30]. However, the problem of composition has not received a thorough study.

**XAspects**. Shonle et al. [38] present a framework for aspect compilation that allows to combine multiple domain-specific aspect extensions. The framework's composition semantics is to reduce all extensions to a single general-purpose aspect extension (AspectJ). Specifically, given a set of programs written in different aspect extensions, XAspects produces a single program in AspectJ. An aspect extension program is translated to one or more AspectJ aspects. In XAspects, collaboration between the aspect extensions is realized as a collaboration between the translated AspectJ's aspects.

The XAspects framework uses a translation-based approach. Specifically, XAspects translates programs in domain-specific aspect extensions to AspectJ. Unfortunately, in the presence of other aspects, this approach does not preserve the behavior of the domain specific aspects, and therefore the XAspects approach does not guarantee a correct result.

Moreover, extensions in XAspects must be reducible to AspectJ. Since only a subset of aspect extensions is expressible in AspectJ, XAspects doesn't achieve composition in general. Our approach to composition and collaboration is not based on translation. In comparison to XAspects our proposed approach is more general.

**Concern Manipulation Environment**. IBM's new Concern Manipulation Environment provides developers with an extensible platform for concern separation: "The CME provides a common platform in which different AOSD tools can interoperate and integrate" [1]. CME would be a natural environment for a large scale application of our approach.

## 7.2 AOP Semantics

Existing works in AOP semantics explain existing aspect extensions and model AOP in general. We base some of our work on these studies. Unfortunately, they do not address the problem of aspect mechanism composition directly.

**Semantics for Existing AOP Languages**. Wand et al.'s [49] semantic for advice and dynamic join points explains a simplified dynamic AspectJ. It provides denotational semantics to a small procedural language, similar to ours. The language embodies key features of dynamic join points, pointcuts and advice. The semantics given does not express the AOP semantics separately from the base. However, advice weaving is defined there as a procedure transformer. This is a special case of a language semantics transformer as we choose to define an aspect mechanism.

Method-Call Interception [27] is another semantical model that gives semantics of advising method calls. Similar to the previously discussed work, it highlights a very specific piece of AOP expressiveness (similar to AspectJ).

**Semantical Models of AOP**. Several studies on AOP semantics provide a general model of AOP functionality. Walker et al. [44] defined aspects through explicitly labeled program points and first-class dynamic advice. Jagadeesan et al. [24] use similar abstrac-tions (pointcuts and advice). Clifton et al. [10, 11] provides parameterized aspect calculus for modeling AOP semantics. In their model, AOP functionality can be applied to any reduction step in a base language semantics. This is similar to the definition of an aspect mechanism we use.

In comparison to our semantics, these models define AOP functionality using low-level language semantics abstractions. Using these more formal approaches for describing our method is left for future work.

**Modular Semantics for AOP**. We define an aspect mechanism separately from the base language and require it to specify only the AOP transformation functionality. This approach leads to the construction of modular AOP semantics. Exploring the application of other approaches for modular language semantics (e.g., modular SOS [31] and monad-based denotational semantics) to describing aspect mechanism is another area for further research.

## 8. CONCLUSION

In this paper we address the open problem of integrating a base language Base with a set of third-party aspect extensions $\text{Ext}_1, \ldots, \text{Ext}_n$ for that language. We present a semantical framework in which independently developed, dynamic aspect mechanisms can be subject to third-party composition and work collaboratively.

We instantiate our approach for aspect mechanisms defined as expression evaluation transformers. The mechanisms can be composed like mixin layers [39, 34, 35] in a pipe-and-filter architecture with delegation semantics. Each mechanism collaborates by *delegating* or *exposing* the evaluation of expressions. The base mechanism serves as the terminator and does not delegate the evaluation further.

We applied our approach in the implementation of a concrete base language MyBase and two concrete aspect extensions to that language: $\text{HisExt}_1$ and $\text{HerExt}_2$. The implementation illustrates the constructions steps. It demonstrates the semantics for third-party composition of aspect mechanisms.

The semantics for $\text{HisExt}_1$ resembles that for AspectJ. Indeed, our approach can be applied to implementing the pointcut and advice mechanism of AspectJ as an aspect extensions to Java. Moreover, our approach is not limited to the pipe-and-filter composition architecture. Introduction of a generalized aspect mechanism model would enable sophisticated compositions of third-party aspect mechanisms. This would further provide a practical way to compose AspectJ with new domain-specific aspect extensions.